\documentclass[10pt, conference, compsocconf]{IEEEtran}
\IEEEoverridecommandlockouts
\usepackage{bm, amsmath,amssymb,graphicx}
\usepackage{subfigure}
\usepackage{algorithm, algpseudocode, algorithmicx}

\newcommand{\mat}[1]{\ensuremath{\bm{\mathrm{#1}}}}
\newcommand{\A}{\ensuremath{\mat{A}}}

\newcommand{\x}{\ensuremath{\mat{x}}}
\newcommand{\X}{\ensuremath{\mat{X}}}

\newcommand{\e}{\ensuremath{\mat{e}}}

\newcommand{\y}{\ensuremath{\mat{y}}}
\newcommand{\Y}{\ensuremath{\mat{Y}}}

\newcommand{\Ps}{\ensuremath{\bm{\Psi}}}

\newcommand{\Ph}{\ensuremath{\mat{\Phi}}}

\newcommand{\trasp}[1]{\ensuremath{#1 ^\mathsf{T}}}

\newcommand{\vect}[1]{\ensuremath{\mathrm{vec}\left \{ #1\right\}}}

\newcommand{\N}{\mathcal{N}}

\def\df{\triangleq}
\def\Ri{\mathbb{R}}

\newcommand{\lzeronorm}[1]{\ensuremath{\left\| #1\right\|_{\ell_0}}}
\newcommand{\lonenorm}[1]{\ensuremath{\left\| #1\right\|_{\ell_1}}}

\newtheorem{remark}{Remark}[section]

\begin{document}

\title{A Novel Progressive Image Scanning and Reconstruction Scheme based on Compressed Sensing and Linear Prediction\vspace{-0.3cm}}

%
\author{\IEEEauthorblockN{Giulio Coluccia, Enrico Magli}
\IEEEauthorblockA{Politecnico di Torino\\
Dipartimento di Elettronica e Telecomunicazioni\\
Torino, Italy\\
\{giulio.coluccia, enrico.magli\}@polito.it
\thanks{The research leading to these results has received funding from the European Research Council under the European Community's Seventh Framework Programme (FP7/2007-2013) / ERC Grant agreement n° 279848}\vspace{-0.3cm}}
}

\maketitle

\begin{abstract}
Compressed sensing (CS) is an innovative technique allowing to represent signals through a small number of their linear projections. In this paper we address the application of CS to the scenario of progressive acquisition of 2D visual signals in a line-by-line fashion. This is an important setting which encompasses diverse systems such as flatbed scanners and remote sensing imagers. The use of CS in such setting raises the problem of reconstructing a very high number of samples, as are contained in an image, from their linear projections. Conventional reconstruction algorithms, whose complexity is cubic in the number of samples, are computationally intractable. In this paper we develop an iterative reconstruction algorithm that reconstructs an image by iteratively estimating a row, and correlating adjacent rows by means of linear prediction. We develop suitable predictors and test the proposed algorithm in the context of flatbed scanners and remote sensing imaging systems. We show that this approach can significantly improve the results of separate reconstruction of each row, providing very good reconstruction quality with reasonable complexity.
\end{abstract}
\vspace{-0.5cm}
\section{Introduction}\label{sec:intro}
\vspace{-0.1cm}

Compressed Sensing (CS) \cite{candes2006cs, donoho2006cs} has recently emerged as an efficient technique for sampling a signal with fewer coefficients than the number dictated by classical Shannon/Nyquist theory. The assumption underlying this approach is that the signal to be sampled is sparse or at least ``compressible'', i.e., it must have a concise representation in a convenient basis. In CS, sampling is performed by taking a number of linear projections of the signal onto pseudorandom sequences. Therefore, the acquisition presents appealing properties such as low encoding complexity, since the basis in which the signal is sparse does not need to be computed, and universality, since the sensing is blind to the source distribution. Reconstruction of a signal from its projections can be done e.g. using linear programming \cite{donoho2006cs}, with a complexity that is $O(N^3)$, with $N$ the number of samples to be recovered. 

Recently, CS has been applied to multi-dimensional signals. In this case, a serious problem arises regarding the computational complexity of the reconstruction process. The conventional approach of measuring the signal along all dimentions at once leads to very large $N$, making the reconstruction computationally intractable. The authors of \cite{duarte2010kronecker} showed a way to recast a multidimensional CS problem to a one-dimensional one, by the means of Kronecker products of sensing and sparsity matrices. 
Reconstruction algorithms for multidimensional signals have also been proposed in \cite{abrardo2011compressive,fowler_mv,wakin_mv} for hyperspectral images and multiview video.

In this paper, we consider the application of CS to devices that acquire 2D visual information through progressive scanning. These devices are equipped with a one-dimensional array of detectors, and a 2D image is obtained via the repeated use of the array over different slices of the 2D object to be imaged. This is a very important scenario, which encompasses many applications. Amongst others, it is worth mentioning at least two examples, which we will focus on in the remainder of this paper. The first is given by {\em flatbed scanners}, where each line of the image is acquired by a 1D optical sensor moving in the orthogonal direction. The second one is represented by {\em airborne and spaceborne imagers} of the pushbroom type for remote sensing applications. In this case, the 1D sensor is carried on a flying platform such as an airplane or satellite; the sensor looks down at the Earth, and acquires a line-by-line scan of the underlying scene, while each line is oriented in the across-track direction, and the platform flight moves the sensor from one line to the next one. These applications, as well as several other ones, can clearly benefit from CS. CS-based imaging systems have been demonstrated in \cite{wakin2006aci}, where the optical sensor is replaced by an array of micromirrors that acquire linear projections of the signal samples via a single detector (\emph{Single-Pixel Camera}). A similar principle can be applied to progressive scanning, where a 1D micromirror array can be used to directly sense lines in the CS format. In the case of the remote sensing imaging system, CS can lead to a simpler and cheaper system, which uses a single detector and produces a reduced number of sampling. Detectors can be costly in the wavelengths outside the visible spectrum, and the reduced number of samples allows to implement simpler onboard processing systems. For the flatbed scanner, CS would be extremely useful in order to develop a scanner of small size, as the CS sensor needs not be of the same physical size as the document being scanned. Moreover, in both cases, processing and data handling would be greatly reduced, which is important in order to reduce power consumption in the remote sensing case, and in order to enable application to small-sized low-power devices in the flatbed scanner case.

In this paper we address this scenario, and tackle the reconstruction problem for 2D images acquired line-by-line. In particular, we propose a simple progressive acquisition algorithm for 2D signals, where rows are acquired independently of each other, but the reconstruction is performed jointly over all rows. Joint reconstruction is achieved through an iterative algorithm that correlates different rows through linear prediction, instead of taking a multidimensional transform as sparsity domain. Prediction allows to exploit correlation in both horizontal and vertical dimensions, even if the acquisition is performed in one direction only. The main concept is to exploit correlation along the vertical direction by iteratively predicting each line and reconstructing the prediction error only, which is sparser than the line itself. Results show that few iterations of the proposed algorithm suffice to significantly improve the MSE of the reconstruction, allowing
to obtain high-quality reconstruction results with feasible complexity.

\vspace{-0.2cm}
\section{Background}\label{sec:background}
\vspace{-0.1cm}

\subsection{Notation and definitions}\label{sec:notation}
\vspace{-0.1cm}

We denote (column-) vectors and matrices by lowercase and uppercase boldface
characters, respectively. The $(m,n)$-th element of a matrix $\A$ is $(\A)_{m,n}$. The $m$-th row of matrix $\A$ is $(\A)_m$. The $n$-th element of a vector $\mat{v}$ is $(\mat{v})_n$. The transpose of a matrix $\A$ is $\trasp{\A}$. The stack operator $\vect{\A}$ denotes the column vector obtained by stacking the columns of \A\ on top of each other, from left to right.

The notation $\lzeronorm{\mat{v}}$ denotes the number of nonzero elements of vector
$\mat{v}$. The notation $\lonenorm{\mat{v}}$ denotes the $\ell_1$-norm of the vector $\mat{v}$
and is defined as $\lonenorm{\mat{v}} \df \sum_i \left |(\mat{v})_i\right |$~. 
 The notation $a\sim\N(\mu,\sigma^2)$ means denotes a Gaussian random variable $a$ with mean $\mu$ and variance $\sigma^2$~.

\vspace{-0.2cm}
\subsection{Compressed Sensing}\label{sec:CS}
\vspace{-0.1cm}

In the standard CS framework, introduced in \cite{candes2006nos}, a signal $\x\in\Ri^{N\times 1}$
 which has  a sparse representation in some basis $\Ps\in\Ri^{N\times N}$, \textit{i.e}:
\begin{equation*}
\x = \Ps \bm{\theta},\quad \lzeronorm{\bm{\theta}} = K,\quad K\ll N
\end{equation*}
can be recovered by a smaller vector $\y\in\Ri^{M\times 1}$, $K<M<N$, of linear measurements $\y = \Ph\x$, where $\Ph\in\Ri^{M\times N}$ is the \emph{sensing matrix}. The optimum solution, requiring at least $M = K+1$ measurements, would be
$$
\widehat{\bm{\theta}}=\arg\min_{\bm{\theta}}\lzeronorm{\bm{\theta}}\ \quad \text{s.t.}\quad \Ph\Ps\bm{\theta} = \y~.
$$
Since the $\ell_0$ norm minimization is a NP-hard problem,
one can resort to a linear programming reconstruction by minimizing
the $\ell_1$ norm
\begin{equation}\label{eq:CS_recovery}
\widehat{\bm{\theta}}=\arg\min_{\bm{\theta}}\lonenorm{\bm{\theta}}\ \quad \text{s.t.}\quad \Ph\Ps\bm{\theta} = \y~,
\end{equation}
provided that $M$ is large enough ($\sim K\log(N/K)$).

The same algorithm holds for signals which are not exactly sparse, but rather compressible, meaning that they (or their representation $\bm{\theta}$ in basis \Ps) can be expressed only by $K$ significant coefficients, while the remaining ones are (close to) zero.

It has been shown in \cite{baraniuk2008spr} that extracting the elements of \Ph\ at random from a Gaussian or Rademacher distribution (i.e., $\pm 1$ with the same probability), and, in general, from any Sub-Gaussian distribution, allows a correct reconstruction with overwhelming probability.

\vspace{-0.1cm}
\section{Proposed Algorithm}\label{sec:prop_alg}
\vspace{-0.1cm}

According to typical progressive scanning approaches, like the ones used by commercial flatbed scanners or by remote sensing systems acquiring environmental pictures, an image is acquired by sensing $N_{\mathsf{COL}}$ pixels of each row in a progressive fashion, until $N_{\mathsf{ROW}}$ rows are acquired. Hence, the acquired image will result as a matrix of pixels of size $N_{\mathsf{ROW}} \times N_{\mathsf{COL}}$, which will be compressed (and, accordingly, decoded) using a conventional technique. This process requires the acquisition (and processing) of $N_{\mathsf{ROW}} N_{\mathsf{COL}}$ pixels. When $N_{\mathsf{ROW}}$ and  $N_{\mathsf{COL}}$ are large, processing of this huge amount of data may represent an issue, especially when dealing with low cost or low complexity devices.

For this reason, we propose a very simple acquisition scheme, based on CS linear measurements taken on each row, without any further processing. This reduces the amount of data to be acquired and processed. The reconstruction algorithm relies on linear predictors in order to  improve the quality of CS reconstruction, by correlating the measurements of adjacent rows in order to exploit their statistical dependencies during the reconstruction stage, largely improving over individual separate reconstruction. The stronger the correlation of pixels within a row and among rows, the better will be the performance of CS reconstruction and of the linear predictor (and hence of the whole reconstruction algorithm).

\vspace{-0.15cm}
\subsection{Image Acquisition}\label{sec:coding}
\vspace{-0.1cm}

The image acquisition algorithm we propose, labelled as Algorithm~\ref{alg:coding}, is very simple and consists in taking linear measurements of each row of the image in a progressive fashion. To minimize the risks of failures in the reconstruction side, a different sensing matrix \Ph\ is drawn for each row.

The image to be measured can be divided into $N_{\mathsf{ROW}}$ rows. For each row, $M$ linear measurements are taken, where $M < N_{\mathsf{COL}}$ and $N_{\mathsf{COL}}$ is the desired vertical resolution. 

In summary, the scene we wish to acquire is represented by the matrix  $\X \in \Ri^{N_{\mathsf{ROW}} \times N_{\mathsf{COL}}}$. For each row of \X, we draw a matrix $\Ph^i \in \Ri^{M \times N_{\mathsf{COL}}}$ whose elements are Gaussian i.i.d. such that $(\Ph^i)_{kj}\sim\N(0,1/M)$, with $k = 1, \ldots, M$ and $j = 1, \ldots, N_{\mathsf{COL}}$. Then, we take $M$ linear measurements of $(\X)_i$ which will form the rows of the matrix of measurements $\Y \in \Ri^{N_{\mathsf{ROW}} \times M}$, namely
$$
\trasp{(\Y)_i} = \Ph^i\trasp{(\X)_i}
$$

\begin{algorithm}[t]
\caption{Proposed acquisition algorithm}\label{alg:coding}
\begin{algorithmic}[1]
\Require the image \X, $M$
\Ensure the measurement matrix \Y
\For{$i = 1$ to $N_{\mathsf{ROW}}$}
\State Draw $\Ph^i$ s.t. $(\Ph^i)_{kj}\sim\N(0,1/M)$
\State$ \trasp{(\Y)_i} \gets \Ph^i\trasp{(\X)_i}$
\EndFor
\State \Return \Y
\end{algorithmic}
\end{algorithm}

A more complex algorithm, based on Compressed Sensing and able to capture spatial correlation in both directions (horizontal and vertical), could acquire in a single shot the whole image in a single measurement vector of length $M'$.
$$
\y' = \Ph'\vect{\X}~,
$$
where $\vect{\X} \in \Ri^{N_{\mathsf{ROW}}N_{\mathsf{COL}} \times 1}$, $\Ph' \in \Ri ^{M' \times N_{\mathsf{ROW}}N_{\mathsf{COL}}}$, $\y' \in \Ri^{M'\times 1}~.$

Even if this algorithm performed better than the one proposed here since the reconstruction would optimally exploit the correlation in 2 dimensions through a 2D transform matrix, it would require the solution of \eqref{eq:CS_recovery} for a vector of length $N=N_{\mathsf{ROW}}N_{\mathsf{COL}}$. For realistic values of $N_{\mathsf{ROW}}$ and $N_{\mathsf{COL}}$, the solution of \eqref{eq:CS_recovery} would be impossible to perform in reasonable time.
On the other hand, the proposed approach splits the problem into smaller (and hence tractable) subproblems. However, in doing so, it does not neglect the spatial correlation in vertical direction, which is modeled and employed in the reconstruction process through the use of linear predictors.

\vspace{-0.2cm}
\subsection{Image Reconstruction}\label{sec:decoding}
\vspace{-0.1cm}

A trivial reconstruction algorithm based on the acquisition scheme described in section \ref{sec:coding} would simply apply the $\ell_1$ reconstruction \eqref{eq:CS_recovery} to recover separately each line of \X\ given the corresponding $\Ph^i$ and $(\Y)_i$.

Instead, we propose an algorithm using this trivial reconstruction as the initialization step and iteratively improves the current estimate of \X\ by modelling statistical dependencies between adjacent lines. We label this Algorithm~\ref{alg:decoding}. We count the iterations using the index $n$. The estimation of \X\ at iteration $n$ is denoted with $\X^{(n)}$.

In particular, the algorithm evaluates a first image reconstruction performing line-by-line separate reconstruction (iteration $n=0$). Then, the iterations start. The intuition is as follows. For each row, if we are able to reliably predict it using the reconstruction of the upper and lower lines at previous iteration with some linear predictor $\mathsf{P(\cdot,\cdot)}$, obtaining $\x_\mathsf{P}$, we can compute the ``measurement'' $\y_\mathsf{P}$ of this prediction by applying matrix $\Ph^i$ to $\x_\mathsf{P}$. Then we calculate the prediction error in the linear measurement domain $\e_{\y}$ by subtracting this ``predicted measurement'' from the original measurement row $(\Y)_i$. The error $\e_{\y}$  will be then reconstructed using \eqref{eq:CS_recovery}, leading to a prediction error on the signal samples equal to $\e_{\x}$. Adding $\e_{\x}$ to $\x_\mathsf{P}$ provides a new estimate of \x. Since the new estimate is more accurate than the old one, the process can be repeated by estimating a new, more accurate predictor.  If the prediction of the row is accurate enough, the prediction error is going to be sparser than the original vector. As a consequence, for an equal number of measurements, the $\ell_1$ reconstruction will yield lower MSE. 

\begin{algorithm}[t]
\caption{Proposed reconstruction algorithm}\label{alg:decoding}
\begin{algorithmic}[1]
\Require the measurement matrix \Y, the set of $\Ph^i$
\Ensure the estimation $\widehat{\X}$
\State $n \gets 0$
\For{$i = 1$ to $N_{\mathsf{ROW}}$}
\State $\widehat{\bm{\theta}} \gets \arg\min_{\bm{\theta}}\lonenorm{\bm{\theta}}\ \quad \text{s.t.}\quad \Ph^i\Ps\bm{\theta} = \trasp{(\Y)_i}$
\State $\trasp{(\X^{(n)})_i} \gets \Ps\widehat{\bm{\theta}}$
\EndFor
\Repeat
\State $n \gets n+1$
\For{$i = 1$ to $N_{\mathsf{ROW}}$}
\If{$i = 1$ or $i = N_{\mathsf{ROW}}$}
\State $\x_\mathsf{P} \gets \trasp{(\X^{(n-1)})_i}$
\Else
\State $\x_\mathsf{P} \gets \trasp{\mathsf{P}\left((\X^{(n-1)})_{i-1},(\X^{(n-1)})_{i+1}\right)}$
\EndIf
\State $\y_\mathsf{P} \gets \Ph^i\x_\mathsf{P}$
\State $\e_{\y} \gets \trasp{(\Y)_i} - \y_\mathsf{P}$
\State $\e_{\bm{\theta}} \gets  \arg\min_{\e}\lonenorm{\e}\ \quad \text{s.t.}\quad \Ph^i\Ps\e = \e_{\y} $
\State $\e_{\x} \gets \Ps\e_{\bm{\theta}}$
\State $\trasp{(\X^{(n)})_i} \gets \trasp{(\x_\mathsf{P} + \e_{\x})}$
\EndFor
\Until{Convergence is reached}
\State \Return $\X^{(n)}$
\end{algorithmic}
\end{algorithm}

In section \ref{sec:num_res_pred}, we test the performance of several linear predictors $\mathsf{P}(\cdot,\cdot)$ and of the overall algorithm. Since \eqref{eq:CS_recovery} is a convex problem and the predictors we test are linear, the overall algorithm can be considered as a projection onto convex sets. This ensures the convergence of the algorithm to the intersection of the constraint sets (if any)  \cite{combettes1993foundations}. 

\begin{remark} We briefly explain here the complexity reduction obtained using Algorithm~\ref{alg:decoding} insted of the standard CS reconstruction algorithm, processing the 2D signal as a whole. For an $N_{\mathsf{ROW}}\times N_{\mathsf{COL}}$ image, the standard CS reconstruction algorithm has an $O(N_{\mathsf{ROW}}^3 N_{\mathsf{COL}}^3)$ complexity. Our algorithm performing $N_{\mathsf{ITER}}$ iterations has an $O(N_{\mathsf{ITER}} N_{\mathsf{ROW}} N_{\mathsf{COL}}^3)$ complexity, with, usually, $N_{\mathsf{ITER}}\ll N_{\mathsf{ROW}}, N_{\mathsf{COL}}$.
\end{remark}

\vspace{-0.2cm}
\section{Numerical Results}
\label{sec:num_res}

\vspace{-0.1cm}
\subsection{Choice of the Predictor}\label{sec:num_res_pred}
\vspace{-0.1cm}

\begin{figure}
\vspace{-0.5cm}
\centering
\includegraphics[width=1\columnwidth]{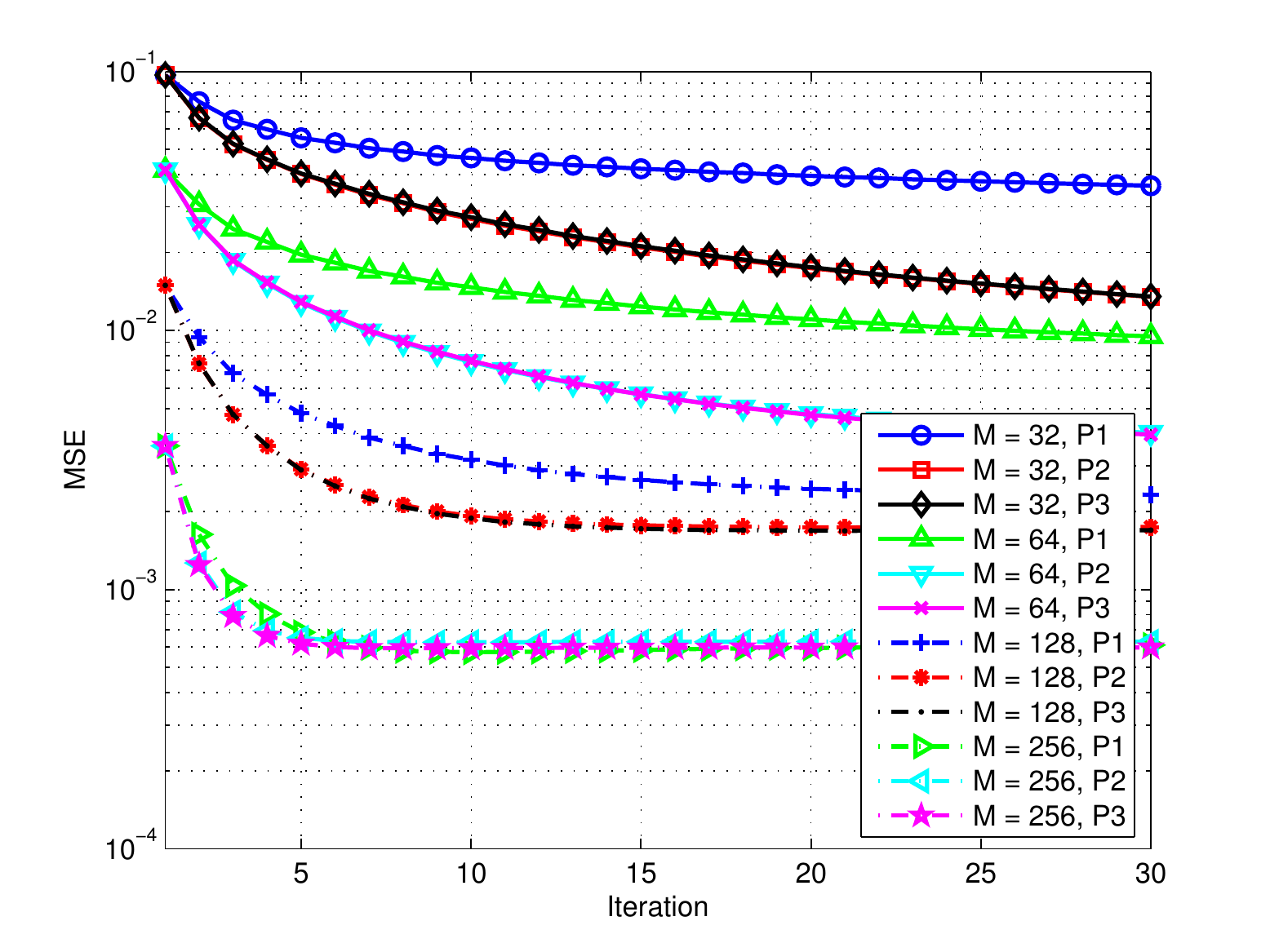}
\vspace{-1cm}
\caption{{Test of different predictors on \emph{lena} $512\times 512$ B/W image}}
\vspace{-0.3cm}
\label{fig:lena}
\end{figure}

First, we start by seeking the linear predictor $\mathsf{P}\left((\X)_{i-1},(\X)_{i+1}\right)$ providing fastest convergence and best MSE performance. For this test, we use the standard \emph{lena} black and white image of size $512 \times 512$. $M$ takes the values of $32, 64, 128, 256$ and the transform matrix $\Ps$ is the DCT matrix. We denote as $\x_\mathsf{P}$ the result of the prediction.

Predictor labelled as \emph{P1} estimates the current\footnote{Here and in the following equations, we omit the index ($n$) denoting current iteration} line to be predicted as the average of the upper and lower lines:
$$
\x_\mathsf{P} = \frac{1}{2}\trasp{\left((\X)_{i-1} + (\X)_{i+1}\right)}
$$

Predictor labelled as \emph{P2} predicts each pixel of current line as the average of adjacent pixels of upper and lower lines
\begin{align}
(\x_\mathsf{P})_j & = \frac{1}{6}\left[(\X)_{i-1,j-i} + (\X)_{i-1,j} + (\X)_{i-1,j+1}\right. \nonumber\\
& + \left.(\X)_{i+1,j-i} + (\X)_{i+1,j} + (\X)_{i+1,j+1}\right]~.\nonumber
\end{align}

Finally, predictor labelled as \emph{P3} predicts each pixel of current line as the \emph{weighted} average of adjacent pixels of upper and lower lines. Weights depend on the distance from the pixel to be predicted, namely
\begin{align}
(\x_\mathsf{P})_j & = \left[a(\X)_{i-1,j-1} + b(\X)_{i-1,j} + a(\X)_{i-1,j+1}\right. \nonumber\\
& + \left.a(\X)_{i+1,j-1} + b(\X)_{i+1,j} + a(\X)_{i+1,j+1}\right]~,\nonumber
\end{align}
with $a = \frac{2-\sqrt{2}}{4}$ and $b = \frac{\sqrt{2}-1}{2}$.

Fig.~\ref{fig:lena} shows the MSE performance of the overall system for different values of $M$ and using the predictors described above. Results show that the convergence is reached for each value of $M$. The bigger $M$, the faster the convergence and the smaller is the MSE at convergence. In any case, it can be noticed that the best performance is obtained for each value of $M$ using predictor labelled as \emph{P3}, i.e. the weighted average. Hence, we will use this predictor in our further tests, omitting to mention it from now on.

For $M=64$, the MSE obtaind with separately recovered lines is $4.16\cdot 10^{-2}$. After 30 iterations, an MSE of $3.96\cdot 10^{-3}$ is obtained, with a gain of 10.2 dB. The convergence in this case is quite slow, but the MSE is decreased as much as one order of magnitude. Faster convergence is obtained with $M=128$, as after 15  MSE is decreased from $1.49\cdot 10^{-2}$ to $1.72\cdot 10^{-3}$, with a gain of 9.38 dB. Finally, with $M=256$ the MSE decreases from $3.59\cdot 10^{-3}$ to $6.18\cdot 10^{-4}$ in 5 iterations only, with a gain of 7.64 dB.

\vspace{-0.2cm}
\subsection{Flatbed Scanner}
\vspace{-0.1cm}

\begin{figure}[t]
\vspace{-0.5cm}
 \centering
 \subfigure[Constellation]
   {\label{fig:constellation}\includegraphics[width=0.4\columnwidth]{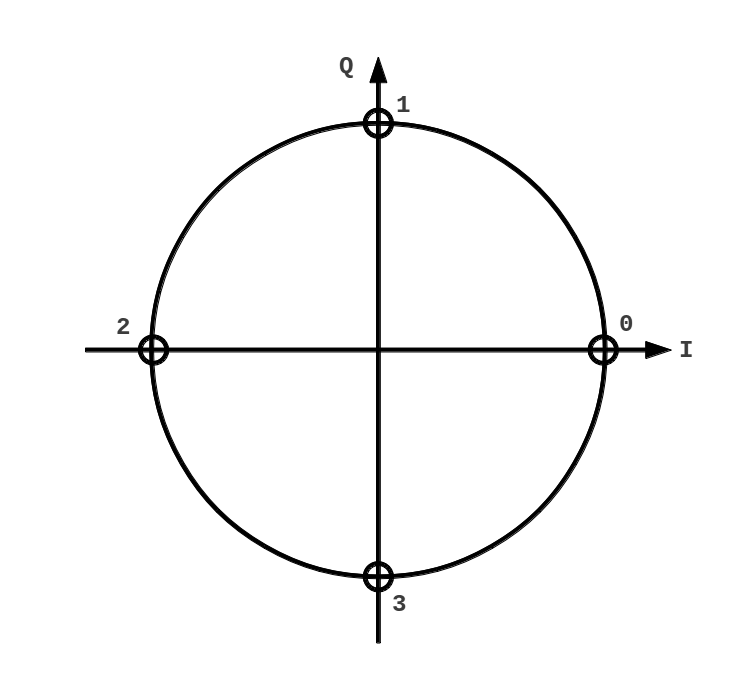}}
 \subfigure[Trellis]
   {\label{fig:trellis}\includegraphics[width=0.4\columnwidth]{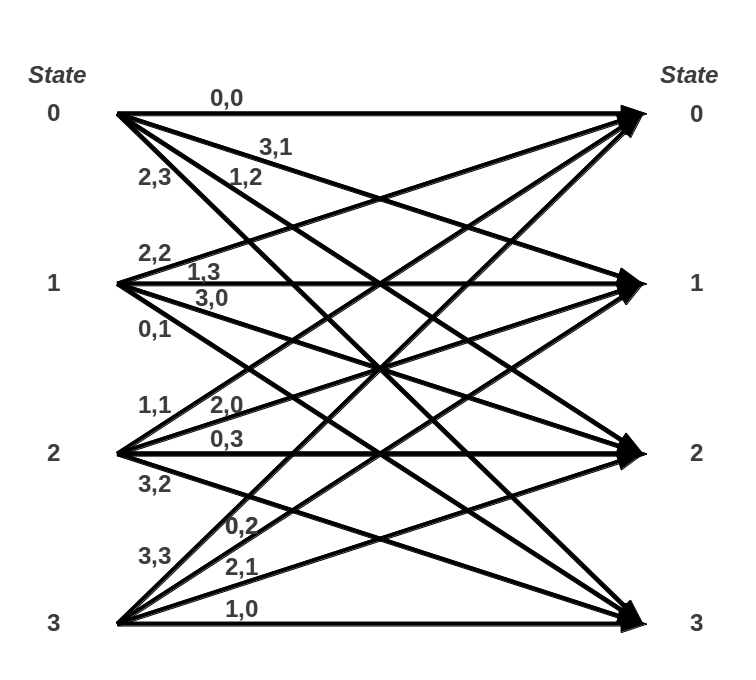}}
\subfigure[Block Diagram]
   {\label{fig:block_diagram}\includegraphics[width=0.8\columnwidth]{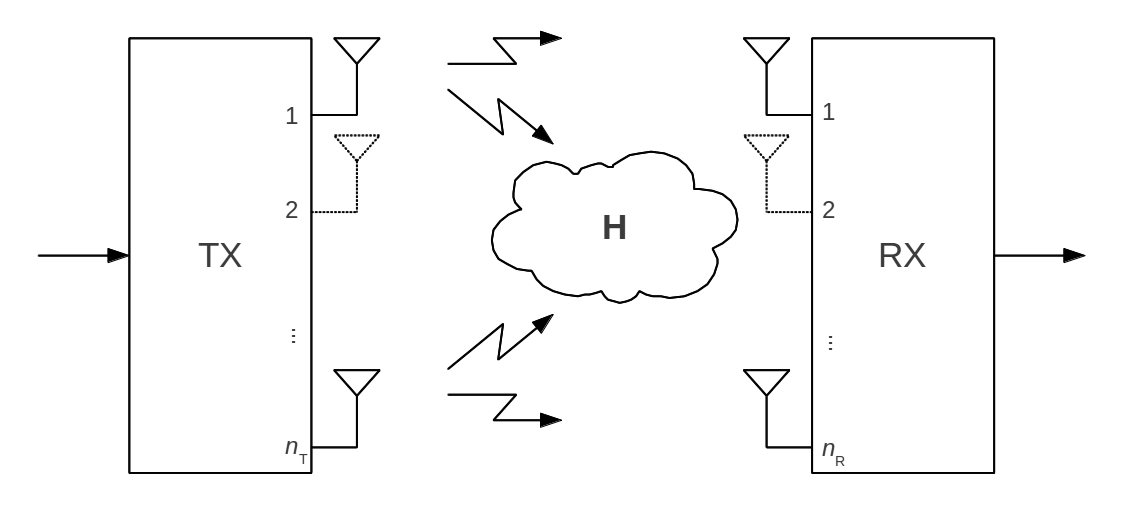}}
   \subfigure[Sample Text]
   {\label{fig:loremIpsum}\includegraphics[width=0.5\columnwidth]{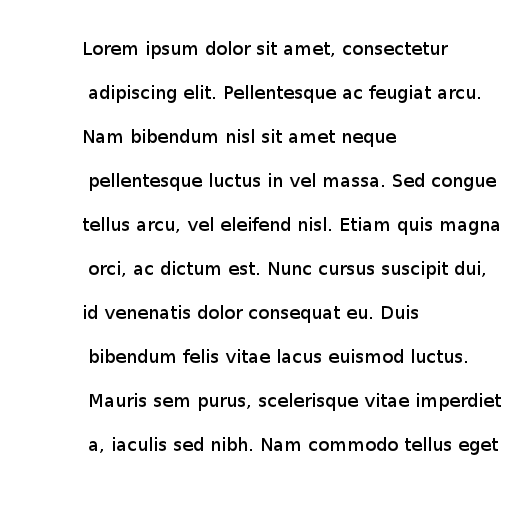}}
\vspace{-0.2cm}
\caption{{The graphics used as test image for \emph{flatbed scanner} scenario}}
\vspace{-0.5cm}
 \label{fig:graphs_bmp}
 \end{figure}

In this section, we apply our algorithm to images suitable to a \emph{flatbed scanner} scenario. These are black and white graphics and text, and are depicted in Fig.~\ref{fig:graphs_bmp}. $M$ takes the values of 8, 16, 32, 64, 128, 256 and (where possible) 512. Since they all have a completely white background (representing paper), they can be considered sparse in the pixel domain. Hence, the matrix \Ps\ is the identity matrix of size $N_{\mathsf{COL}}$, namely $\mat{I}_{N_{\mathsf{COL}}}$.

Fig.~\ref{fig:constellation} is the simplest graphic, representing a QPSK constellation. Fig.~\ref{fig:trellis} represents a slightly more complicated (hence, less sparse) graphic, the trellis of a convolutional code. Fig.~\ref{fig:block_diagram} is a larger figure representing a generic block diagram. Finally, fig.~\ref{fig:loremIpsum} depicts a sample of generic text.

Table~\ref{tab:scanner_res} reports the results obtained using the proposed algorithm. The table shows, for each image, the initial MSE (obtained using separate CS reconstruction of each line), the MSE the algorithm converges to, the performance gain, and the number of iterations necessary to reach convergence. Figures confirm the results obtained in the previous section. The more measurements are taken, the faster is the convergence and the lower is the MSE that can be obtained when the algorithm has converged. When the picture is very sparse, it is possible to obtain a reduction of one order of magnitude, while when the picture is less sparse the contribution of Compressed Sensing is weaker, but still a reduction of about 50\% in MSE can be obtained.

\begin{table}[t]
\caption{{MSE and convergence results on sample graphics}}
\centering
\begin{tabular}{|c|c|c|c|c|}
\hline
$M$ & init. MSE & conv. MSE & gain (dB) & steps \\
\hline
\multicolumn{5}{|c|}{\emph{Constellation} ($680\times576$)} \\
\hline
64 & $2.08\cdot 10^{-2}$ & $7.86\cdot 10^{-3}$ & 4.23 & 18 \\
\hline
128 & $9.99\cdot 10^{-3}$& $3.05\cdot 10^{-3}$ & 5.15 & 14\\
\hline
256 & $4.72\cdot 10^{-3}$ & $7.25\cdot 10^{-4}$ & 8.14 & 10 \\
\hline
\multicolumn{5}{|c|}{\emph{Trellis} ($680\times576$)} \\
\hline
64 & $8.56\cdot 10^{-2}$ & $3.98\cdot 10^{-2}$ & 3.33 & 18 \\
\hline
128 & $7.45\cdot 10^{-2}$& $2.00\cdot 10^{-2}$ & 5.71 & 13\\
\hline
256 & $3.39\cdot 10^{-2}$ & $6.51\cdot 10^{-3}$ & 7.17 & 8 \\
\hline
\multicolumn{5}{|c|}{\emph{Block Diagram} ($529\times1123$)} \\
\hline
64 & $8.38\cdot 10^{-3}$ & $7.02\cdot 10^{-3}$ & 0.77 & 7 \\
\hline
128 & $5.79\cdot 10^{-3}$& $3.97\cdot 10^{-3}$ & 1.64 & 6\\
\hline
256 & $2.79\cdot 10^{-3}$ & $1.66\cdot 10^{-3}$ & 2.25 & 5 \\
\hline
512 & $1.23\cdot 10^{-3}$ & $4.71\cdot 10^{-4}$ & 4.17 & 5 \\
\hline
\multicolumn{5}{|c|}{\emph{Sample Text} ($512\times512$)} \\
\hline
64 & $6.40\cdot 10^{-2}$ & $4.59\cdot 10^{-2}$ & 1.44 & 10 \\
\hline
128 & $5.46\cdot 10^{-2}$& $2.99\cdot 10^{-2}$ & 2.62 & 7\\
\hline
256 & $3.39\cdot 10^{-2}$ & $1.41\cdot 10^{-2}$ & 3.81 & 4 \\
\hline
\end{tabular}
\label{tab:scanner_res}
\end{table}



\vspace{-0.3cm}
\subsection{Remote Sensing Image Acquisition}
\vspace{-0.15cm}

\begin{figure}
 \centering
 \includegraphics[scale=0.9]{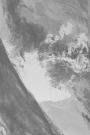}
\vspace{-0.4cm}
 \caption{{The AIRS sensor gran 9 hyperspectral image, $600$-th band}}
 \vspace{-0.5cm}
 \label{fig:airs_bmp}
 \end{figure}

To test the performance of the proposed scheme in a \emph{remote sensing} scenario, we use a spectral band extracted from hyperspectral image ``granule 9'' of the AIRS sensor. AIRS is an ultraspectral sounder with 2378 spectral channels, used to create 3D maps of air and surface temperature. The spatial size is $N_{\mathsf{COL}} = 90$ pixels and $N_{\mathsf{ROW}} = 135$ lines. The dataset consists in the raw output of the detector, without any processing, calibration or denoising applied. We choose the $600$-th band, which is depicted in Fig.~\ref{fig:airs_bmp}, but very similar results have been obtained with other bands and are omitted for brevity. $M$ takes the values of 8, 16, 32 e 64. The sparsity basis \Ps\ is the DCT.

Table~\ref{tab:airs9_600_results} (Basic Algorithm) summarizes the results obtained applyng the proposed algorithm to the $600$-th band of the test image. Results show that with $M=8$ and $M=16$ the convergence is very slow and is not reached after 20 iterations. On the other hand, when $M=32$ the convergence is obtained after 10 iterations (reducing from $2.17\cdot 10^{-2}$ to $1.56\cdot 10^{-3}$, with a gain of 11.4 dB), while taking $M=64$ measurements per row implies the convergence after 4 steps only (with MSE reduction from $2.89\cdot 10^{-3}$ to $3.40\cdot 10^{-4}$ and a gain of 9.29 dB).
\begin{table}[t]
\caption{{MSE and convergence results on AIRS sensor image}}
\centering
\begin{tabular}{|c|c|c|c|c|}
\hline
$M$ & init. MSE & conv. MSE & gain (dB) & steps \\
\hline
\multicolumn{5}{|c|}{Basic algorithm} \\
\hline
8 & $2.40\cdot 10^{-1}$ & $2.63\cdot 10^{-2} $ (20-th it.) & 9.6 & 20+ \\
\hline
16 & $9.57\cdot 10^{-2}$ & $5.21\cdot 10^{-3}$ (20-th it.) & 12.6 & 20+ \\
\hline
32 & $2.17\cdot 10^{-2}$& $1.56\cdot 10^{-3}$ & 11.4 & 10\\
\hline
64 & $2.89\cdot 10^{-3}$ & $3.40\cdot 10^{-4}$ & 9.29 & 4 \\
\hline
\multicolumn{5}{|c|}{Kronecker improved algorithm} \\
\hline
8 & $4.60\cdot 10^{-3}$ & $3.80\cdot 10^{-3}$ & 0.83 & 7 \\
\hline
16 & $2.62\cdot 10^{-3}$ & $2.02\cdot 10^{-3}$ & 1.13 & 5 \\
\hline
32 & $1.22\cdot 10^{-3}$& $9.73\cdot 10^{-4}$ & 0.98 & 3\\
\hline
64 & $2.95\cdot 10^{-4}$ & $2.64\cdot 10^{-4}$ & 0.48 & 1 \\
\hline
\end{tabular}
\label{tab:airs9_600_results}
\end{table} 

\vspace{-0.15cm}
\subsection{Improving performance with Kronecker CS}\label{sec:kcs}
\vspace{-0.15cm}

An improvement to the performance of the algorithm is obtained plugging the so-called \emph{Kronecker Compressed Sensing} (KCS), described in \cite{duarte2010kronecker}, into the algorithm we propose.

KCS allows the acquisition of a multidimensional signal using linear measurement and reconstruction of the entire signal, allowing to exploit sparsity entailed in each signal dimension.

Considering  a 2D \X, it is easy to show that, if $$ \X = \trasp{\Ps_{\mathsf{ROW}}}\mat{\Theta} \Ps_{\mathsf{COL}}~,$$ where $\mat{\Theta}\in \Ri^{N_{\mathsf{ROW}} \times N_{\mathsf{COL}}} $ is the matrix collecting the separable 2D transform coefficients and $\Ps_{\mathsf{ROW}} \in \Ri^{N_{\mathsf{ROW}} \times N_{\mathsf{ROW}}} $ and $\Ps_{\mathsf{COL}} \in \Ri^{N_{\mathsf{COL}} \times N_{\mathsf{COL}}} $ are the matrices representing the basis applied to rows and columns, respectively, then
$$
\vect{\trasp{\X}} = \left(\trasp{\Ps_{\mathsf{ROW}}}\otimes\trasp{\Ps_{\mathsf{COL}}}\right)\vect{\trasp{\mat{\Theta}}}~.
$$

Hence, each multidimensional signal can be reshaped as a column vector (in this case, we consider transposed images because we measure image rows), acquired with a ``global'' sensing matrix $\Ph'$ and the reconstruction problem \eqref{eq:CS_recovery} can be recast to multiple dimensions using Kronecker products.

Two KCS sampling techniques are proposed in \cite{duarte2010kronecker}. The first is similar to the one described in the very last paragraphs of section \ref{sec:coding}, where the sensing matrix $\Ph'$ is dense. As already stated, the drawback of this approach is the infeasible reconstruction complexity in realistic conditions. The second, instead, allows the acquisition of separate measurements of each row, but joint reconstruction through the use of the 2D transform $\trasp{\Ps_{\mathsf{ROW}}}\otimes\trasp{\Ps_{\mathsf{COL}}}$. This means that $\Ph'$ is a \emph{block diagonal} $N_{\mathsf{ROW}}M\times N_{\mathsf{ROW}}N_{\mathsf{COL}}$ matrix, where each of the $N_{\mathsf{ROW}}$ blocks is the $M\times N_{\mathsf{COL}}$ matrix $\Ph^i$ sensing each row. The particular structure of $\Ph'$ allows to solve \eqref{eq:CS_recovery} with reasonable complexity.

Hence, we use KCS to initialize the iterative algorithm we propose in this paper (instead of separate linewise reconstruction) and apply it to the \emph{remote sensing} scenario. The performance of this modified version of the algorithm are reported in Table.~\ref{tab:airs9_600_results} (Kronecker improved algorithm). The figures show two effects. First, the initial MSE is much lower than in the separate reconstruction case. This gain can be noticed in particular when $M$ is small and is due to the better performance of KCS reconstruction with respect to separate reconstruction; second, the iterative algorithm slightly improves the overall performance and converges in very few steps. This is due to the fact that KCS captures also correlation in vertical direction, making the contribution of each iteration less effective.

\begin{figure}
\vspace{-0.5cm}
\centering
\includegraphics[width=1\columnwidth]{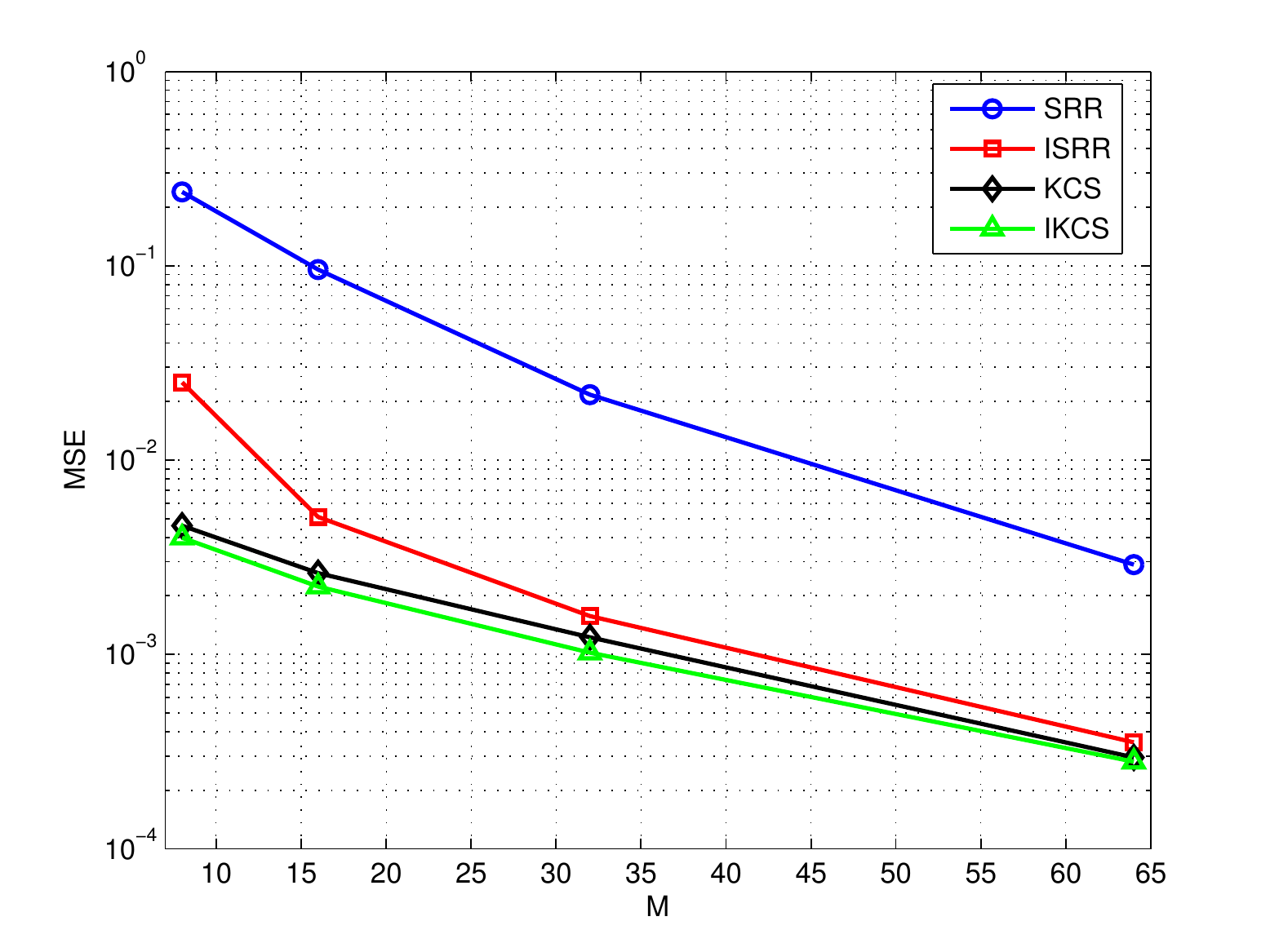}
\vspace{-1cm}
\caption{{Performance comparison of proposed algorithms vs. $M$}}
\vspace{-0.5cm}
\label{fig:overall}
\end{figure}

Figure~\ref{fig:overall} summarizes the best MSE performance obtained by Separate Row Reconstruction (SRR), our Iterative algorithm initialized with Separate Row Reconstruction (ISRR), the Kronecker Compressed Sensing (KCS) and our Iterative algorithm with KCS initialization (IKCS) vs. the number of measurements $M$. Best performing algorithms are the ones implementing KCS. Plain KCS shows a gain of 7.37 dB over ISSR when $M=8$, and 1.10 dB when $M=32$. When using IKCS, roughly 1 dB of additional gain can be obtained with very few iterations.


Finally, as a term of comparison, we report here the MSE performance of a simple reconstruction algorithm named \emph{Orthogonal Matching Pursuit} \cite{tropp2007signal}, whose complexity is linear in the number of samples of the original signal ($N_\mathsf{ROW}N_\mathsf{COL}$ in this case). We acquire and reconstruct the entire image as a whole using $M=32\cdot N_\mathsf{ROW}$ and $M=64\cdot N_\mathsf{ROW}$ measurement, to be compared with the performance of our algorithm with $M=32$ and $M=64$, respectively. For $M=32\cdot N_\mathsf{ROW}$, we obtain an MSE of $1.9\cdot 10^{-3}$, while for $M=64\cdot N_\mathsf{ROW}$ we obtain an MSE of $1.7\cdot 10^{-3}$. Hence, our algorithm with $M=32$ performs 3 dB better than OMP with the same total amount of measurements, while with $M=64$ the gain is 8 dB.

\vspace{-0.3cm}
\section{Conclusions}\label{sec:conclutions}
\vspace{-0.1cm}

In this paper, we proposed a simple and effective algorithm to acquire a 2D signal entailing correlation in both horizontal and vertical directions, like an image. The acquisition is performed by separately acquire each line of the image, taking a number of random linear projections smaller than the number of pixels composing the row itself, as in the Compressed Sensing paradigm. The reconstruction process consists of an iterative algorithm based on the linear prediction of a line and the CS reconstruction of the prediction error, which is supposed to be sparser than the original vector.

We apply this algorithm to two scenarios: flatbed scanners and remote sensing applications. We show that applying our algorithm to images typical of these scenarios it is possible to improve in few iterations the quality obtained reconstructing each row separately.

\vspace{-0.1cm}

\end{document}